%% file: 01_main.tex
\documentclass[sigconf]{acmart}
\usepackage{xspace}
\newcommand{\ie}{\emph{i.e.,}\xspace}
\newcommand{\eg}{\emph{e.g.,}\xspace}
\newcommand{\Eg}{\emph{E.g.,}\xspace}

\newcommand{\etal}{\emph{et~al.}\xspace}

\setcopyright{acmlicensed}
\copyrightyear{2026}
\acmYear{2026}
\acmDOI{XXXXXXX.XXXXXXX}
\acmConference[GREENS '26]{10th International Workshop on Green and Sustainable Software}{Tuesday, April 14 2026}{Rio de Janeiro, Brazil}
\acmISBN{978-1-4503-XXXX-X/2026/06}

\usepackage{cleveref}
\usepackage{enumitem}
\usepackage{booktabs}
\usepackage[table]{xcolor}
\usepackage{tabularx}

\usepackage[most]{tcolorbox}

\usepackage[labeled,resetlabels]{multibib}
\newcites{P}{Primary Studies}

\usepackage[subtle,title=normal,tracking=normal]{savetrees}

\begin{document}

\title{Injecting Sustainability in Software Architecture: A Rapid Review}

\author{Markus Funke}
\email{m.t.funke@vu.nl}
\orcid{0000-0003-2302-2555}
\affiliation{%
  \institution{Vrije Universiteit Amsterdam, The Netherlands}
  \country{}
}

\author{Patricia Lago}
\email{p.lago@vu.nl}
\orcid{0000-0002-2234-0845}
\affiliation{%
  \institution{Vrije Universiteit Amsterdam, The Netherlands}
  \country{}
}

\renewcommand{\shortauthors}{Funke and Lago}


\begin{abstract}
Sustainability has evolved from an emerging concern into a fundamental responsibility in software design, development, and operation. Research increasingly explores how sustainability can be systematically integrated into existing software engineering practices. Building on an industry–academia collaboration, we contribute to this discourse by conducting a mixed-method empirical study. We combine a rapid review of secondary studies with a focus group of practitioners. The review identifies challenges and opportunities in embedding sustainability in software architecture, while the focus group enriches and compares these findings. Based on the literature and industry synthesis, we derive five tangible takeaways to inform architects working in the field, and to guide our industry partners in the integration of sustainability concerns in architecture practices.
\end{abstract}





\maketitle

\input{02_content}

\begin{acks}
Part of this research is funded by the Sustainable IT - Lab of the Vrije Universiteit Amsterdam and ABN AMRO Bank N.V. We thank the participating experts and architects. In preparation of this work the authors used \textit{Writefull} to improve readability and language. After using this tool, the authors reviewed and edited the content as needed and take full responsibility for the content of the publication.
\end{acks}

\bibliographystyle{ACM-Reference-Format}
\bibliography{03_references}

\bibliographystyleP{ACM-Reference-Format}
\bibliographyP{04_referencesP}

\end{document}

%% file: 02_content.tex
\section{Introduction}
\label{sec:introduction}
The environmental footprint of software systems, such as those based on cloud computing or artificial intelligence, already compares to that of sectors traditionally associated with high emissions and continues to grow \cite{WorldEconomicForum_2025, MalmodinEtAl_ICTSector_2024}. With the rapid expansion of data-intensive technologies, the consequences for our environment and society have become impossible to overlook \cite{bessin2023news, manner2023black, DesislavovEtAl_TrendsAI_2023}. As a result, sustainability has evolved from an emerging concern into a fundamental responsibility in software development and operation.

Sustainability as a research field has reached a sufficient maturity-level since its earliest explorations around the early 2000s \cite{CaleroEtAl_5WsGreen_2020} and can now be considered as a ``stable line of research'' \cite{CaleroEtAl_5WsGreen_2020}. Over the past years, research also acknowledges the multi-dimensional nature of sustainability in the context of software-intensive systems, recognizing the dimensions of focus (\ie environmental, technical, social, and economic) and time (\ie direct, indirect, systemic) \cite{LagoEtAl_Toolkit_2024}.

Meanwhile, the software industry has similarly recognized sustainability as a relevant and pressing concern \cite{HeldalEtAl_SustainabilityCompetencies_2024}. Also, research efforts with industry increasingly address how sustainability concerns can be systematically integrated into existing software engineering practices, revolving around concerns such as tackling it from the knowledge perspective \cite{FunkeEtAl_ApproachCarve_2025}, or from educational and training perspective \cite{PetersEtAl_SustainabilityComputing_2024}. Also in our own collaborations with experts in the field and organizations, similar questions emerge regarding how sustainability concerns can be operationalized and addressed in the current way of working. 
Yet, most work approaches this challenge from a \textit{theoretical}, general \textit{software engineering} viewpoint, leaving a \textbf{gap} for empirically grounded reflections tailored to \textit{software architecture practice}.

The authors of this paper are members of a national working group\footnote{(in Dutch) \url{https://coalitieduurzamedigitalisering.nl/werkgroepen/}} in the Netherlands on digital sustainability in software architecture, which brings together practicing architects and researchers. The group aims to strengthen collaboration between research and industry. The \textbf{goal} of this group is (among others) to \textit{understand how sustainability can be integrated into the software architecture process and to provide architects with concrete guidelines and future working directions}. 
This goal serves as the underlying objective for our study. We aim to contribute empirical insights as a starting point for future collaboration and further work within the group. For our investigation, we define the following overarching research question (RQ):

\begin{enumerate}
    \item[\textbf{RQ}] What are the challenges related to integrating sustainability into software architecture?
\end{enumerate}

To address this question, we build on the industry–academia collaboration established within the working group and conduct a mixed-method empirical study. First, we analyze the scientific literature through a (meta-)rapid review. Second, we compare these findings in a focus group with 11 experts working in the field.

\section{Related Work}
\label{sec:relatedwork}

\textit{Literature Reviews.}
A number of studies have conducted literature reviews in the intersection of sustainability and software engineering. However, we found only a few tertiary studies which relate directly to our work. García-Mireles \etal \cite{Garcia-MirelesEtAl_SustainabilityField_2025} analyze 80 systematic reviews on sustainability in software engineering. Although the study highlights the need for stronger research-industry collaboration, its main goal is to synthesize implications and opportunities for research and to provide a roadmap for academia.

Similar to our work, the tertiary study from Gross and Ouhbi~\cite{GrossOuhbi_ClearingPath_2024} identifies challenges, though again from a general software engineering perspective rather than through the lens of software architecture. Based on 18 secondary studies, they distill key challenges and propose five recommendations, including the call for more actionable research.
Other tertiary reviews address more specific intersections such as sustainability in Internet of Things and Smart Cities \cite{MottaEtAl_IntersectionInternet_2024} or blockchain \cite{Schinckus_GoodBad_2020}.

Beyond tertiary studies, other literature reviews exist analyzing sustainability in different contexts. For example, Calero \etal \cite{CaleroEtAl_5WsGreen_2020} examine 542 studies to map the research field, concluding that sustainable software research has reached a mature stage, without proposing practical implications. Trinh \etal \cite{TrinhEtAl_SustainabilityIntegration_2024} explore how AI integration in the software development lifecycle affects sustainability, while Peters \etal \cite{PetersEtAl_SustainabilityComputing_2024} focus on sustainability in computing education.

\textit{Rapid Reviews.}
We identified only two rapid reviews connecting sustainability with software engineering practice. Kumar \etal~\cite{KumarEtAl_BalancingProgress_2024} partner with a financial institution to analyse sustainability benefits and costs of AI systems while enriching their results with interviews. Xiao~\cite{Xiao_ArchitecturalTactics_2024} conducts a similar study to synthesize architectural tactics for sustainable microservices.

\textit{Industry Studies.}
Some other studies approach the topic directly in industry. Heldal \etal~\cite{HeldalEtAl_SustainabilityCompetencies_2024}, for instance, investigate sustainability competencies among practitioners through interviews and focus groups to identify needs for education and training. They conclude that companies are indeed motivated to address sustainability concerns within their business, while the experts are unclear about the foundational concepts of sustainability. Lammert \etal~\cite{LammertEtAl_SoftwareEngineers_2022} on the other hand, tackle sustainability from the requirements engineering and sociological angle. They interview software engineers, and conclude that experts should be more involved in the design process of the actual software. Also other studies (\eg \cite{FunkeEtAl_ApproachCarve_2025, KaritaEtAl_SoftwareIndustry_2021, GroherWeinreich_InterviewStudy_2017}) executed together with industry suggest that sustainability should be addressed at architecture level.

\textit{Summary.}
In contrast to the studies discussed above, our research wants to bridge the gap between literature and industry by identifying a set of reflections specifically tailored to software architecture practice. Additionally, our study is directly motivated by software architects working in the field. Rather than taking a general software engineering perspective, we adopt a software architecture focus.

\section{Method}
\label{sec:method}
\begin{figure}[h]
  \centering
  \includegraphics[width=1\columnwidth]{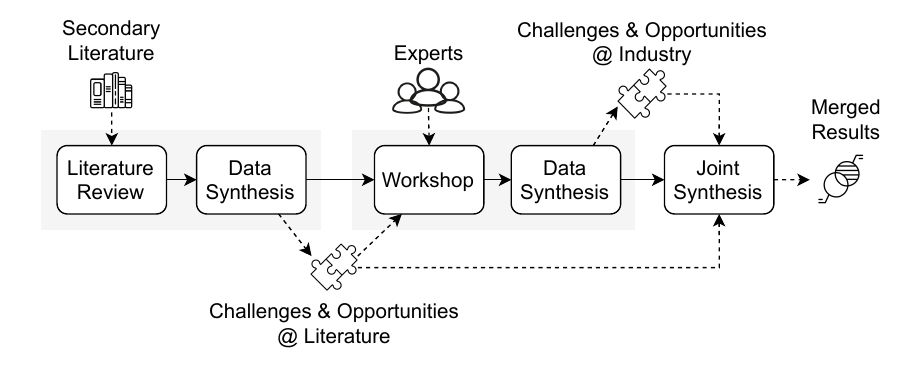}
  \caption{General study design}
  \label{fig:study-design}
  \Description{General study design.}
\end{figure}

To answer our RQ, we design a mixed-method research, as illustrated in our study design in \Cref{fig:study-design}. The literature review aims to analyze the scientific state of the art regarding challenges and opportunities while integrating and addressing sustainability concerns within software architecture practices. We then conduct a focus group with professionals working in different domains, addressing two goals: (i) exploring their own challenges and opportunities, and (ii) comparing the findings from the literature with industry experience. Both results are then analyzed together. Based on this synthesis, we frame tangible takeaways to guide our industry partners in the adoption of sustainability-aware architecture practices.
To increase transparency and reproducibility, all relevant data and artifacts are available in an online replication package \cite{ReplicationPackage}.

\subsection{Meta-Rapid Review}
\label{sec:method-RR}
While systematic literature reviews \cite{KitchenhamCharters_GuidelinesPerforming_2007} or systematic mapping studies \cite{PetersenEtAl_SystematicMapping_2008} provide a rigorous means to analyze the scientific state of literature, they are usually time-intensive and often sacrifice the timeliness of their results. As our problem emerged from an industrial context (cf. \Cref{sec:introduction}), timeliness and practical relevance are crucial.

A rapid review \cite{CartaxoEtAl_RapidReviews_2020} addresses these needs by applying a simplified yet systematic secondary study approach. It adapts review procedures to time and effort constraints, trading some rigor for speed and industry relevance. In our case, we further adapt this approach by conducting not a secondary but a \textbf{tertiary review} \cite{KitchenhamCharters_GuidelinesPerforming_2007}, focusing on existing literature reviews in the field of sustainability and software architecture. Therefore, we refer to our method as a \textit{meta-rapid review}. Unlike typical rapid reviews, which are often derived from the needs of a single industrial partner, our problem emerged from a group of architects working in different companies, as mentioned in \Cref{sec:introduction}. To support this effort, our meta-rapid review provides a timely and evidence-based overview of the current state of research. As in rapid reviews usual, the review was mostly performed solely by one author, \ie the first author of this paper. The second author contributed in building the search term, solving uncertainties during the study selection, and discussions for the thematic analysis. Our findings will be reported back to our experts in form of an evidence briefing, which is also part of the replication package online \cite{ReplicationPackage}.

\subsubsection{Search strategy}
Our search string was driven by our RQ and problem derived from industry. After collecting relevant literature and performing pilot queries, we formulated the following query:

\begin{enumerate}
    \item[Q:] \small\texttt{((software AND (architecture OR architectures OR ar\-chi\-tec\-tur\-al)) OR architecting) AND (sus\-tain\-a\-bil\-i\-ty OR sus\-tain\-able) AND (systematic OR literature OR mapping OR review)}
\end{enumerate}

The query was executed using two source-neutral meta search engines for scientific literature: Google Scholar and Scopus. Instead of relying on specific digital libraries such as the ACM Digital Library or IEEE Xplore, we chose these search engines to reduce complexity of our rapid review.

For Scopus, the query was extended with a broader set of synonyms to better capture relevant secondary studies. The query shown in \texttt{Q} was optimized for Google Scholar due to Google Scholar's restriction on the number of search terms. However, our initial search revealed that some relevant studies were missing. To mitigate this and capture secondary studies that do not explicitly label themselves as ``tertiary studies'' but use terms such as ``meta-analysis'' or ``analysis of research'', we included such term variations in the Scopus query.

Both complete search queries are available in our online replication package \cite{ReplicationPackage}. For Google Scholar, we applied the query only to the \texttt{TITLE} field, whereas in Scopus we used the \texttt{TITLE-ABS-KEY} field, covering the study title, abstract, and keywords. To further mitigate missing out on potential studies, we performed one round of forward and backward snowballing \cite{Wohlin_GuidelinesSnowballing_2014} on our selected studies by using the Scopus advanced search feature.

\subsubsection{Study selection}
For Google Scholar we used the tool Publish or Perish\footnote{Publish or Perish - \url{https://harzing.com/resources/publish-or-perish}} to execute the query; for Scopus the web-interface was used. Both offer to export the search results in CSV format which allowed us to use Zotero\footnote{\url{https://www.zotero.org/}} as literature management tool to merge the results, identify duplications, and retrieve corresponding files (\ie PDFs). After the first merge and cleaning phase we moved all results to a spreadsheet to perform the inclusion and exclusion based on the following criteria:

\begin{enumerate}
    \item[I1] Study is a secondary or tertiary study
    \item[I2] Study is on software architecture
    \item[I3] Study is related to at least one sustainability dimension
    \item[E1] Study is not in English
    \item[E2] Study full-text is not accessible
    \item[E3] Study is not peer reviewed
\end{enumerate}

\subsubsection{Data extraction}
During the final study selection phase, we already extracted the most obvious paper meta-data (\textit{type of study} such as SLR or SMS; \textit{paper focus} such as architecture in general, cloud architecture, architecture evaluation; or \textit{target quality concerns} such as maintainability, energy efficiency) as well as the targeted sustainability dimensions (technical, environmental, economic, social, or other). If we discovered a paper as not fitting during this data extraction phase, we excluded it.

\subsubsection{Coding and Synthesis}
For the actual data extraction, we followed a systematic coding and thematic analysis \cite{CruzesDyba_RecommendedSteps_2011} process. First, we used ATLAS.ti\footnote{\url{https://atlasti.com/}} to extract direct paper quotes and code first challenges, opportunities, categories or concepts across the final set of studies (\textit{open coding}). With challenges and opportunities we mean potential implications that still hinder or could enable addressing sustainability concerns into software architecture and related processes and activities. We are not interested in how to improve the sustainability of certain architecture styles (like microservices, cloud, data centers, etc.). We focus solely on the architecture process. All identified quotes and their preliminary codes were then exported from ATLAS.ti and fed back to the working spreadsheet.

We moved then to another iterative coding cycle and translated codes into higher-order themes (\textit{theming}). We were open to let codes and themes change, to get them merged, or to get them split during the entire process. After the last cycle we finished with the final code book (cf. replication package \cite{ReplicationPackage}). The codes and themes were eventually aggregated to 25 sub-categories, classified in 5 high-level categories, with identifying 91 challenges and 68 opportunities. 

\subsection{Focus Group}
\label{sec:method-WS}
To compare and enrich our literature study findings with the perspectives from industry, we conducted an expert focus group in the working group of the national coalition (cf. \Cref{sec:introduction}). Focus groups as qualitative research method are designed to gather data in the form of insights, opinions, and attributes from a group of participants through group interaction \cite{KontioEtAl_FocusGroup_2008}. Typically, such groups are composed of diverse participants (usually between 3-12 \cite{KontioEtAl_FocusGroup_2008}) with different professional backgrounds who discuss a specific topic under the guidance of a moderator. 

\subsubsection{Execution}
To derive our results, we provided the participants with a simple scenario\footnote{Scenario: \textit{``You \textbf{have to} consider sustainability as a requirement in your next design.''}}, physical sticky notes, and asked them what methods, practices, challenges, or opportunities come to their mind while considering the provided scenario. To ensure equal participation in our rather large group, the participants were asked to write 3-5 ideas individually. Each idea should be written on one sticky note, carry their initials (for a potential follow up), and their personal priority for this idea. After 10 minutes, we asked the participants to select their personal top-3 ideas, pitch them to the whole group, and put them on a wall where we collected the results. We especially allowed group discussions during this `pitch' and encouraged the participants to already (loosely) group their results on the wall (if feasible). The session was moderated by the first author, while the second author took notes and raised follow up questions.

\subsubsection{Analysis}
After the session, the collected sticky notes were digitized by transcribing them into digital notes using Miro\footnote{\url{https://miro.com/}}. Both authors then performed a thematic analysis on the digitalized data---independently. After identifying high-level categories individually, the authors achieved consensus by discussing and consolidating their results, merging the 40 notes into five distinct categories.

\subsection{Threats to Validity}
\label{sec:ttv}
Following the recommendations of Verdecchia \etal \cite{VerdecchiaEtAl_ThreatsValidity_2023a}, we do not treat validity considerations as afterthought. Instead, we address them as part of our method, as certain study design trade-offs were already made during our research design. We structure the three most relevant threats according to Wohlin~et~al.~\cite{WohlinEtAl_ExperimentationSoftware_2012}.

\textit{External Validity.}
Even though our literature review followed a systematic process and was validated through a focus group, the generalizability of our results remains limited. While it is likely that we covered most relevant studies using the two meta search engines, yet some may have been missed despite the snowballing. We therefore decided to conduct the focus group, involving 11 experts from 10 organizations across 6 domains, to complement the literature and provide an initial understanding of industry perspectives, rather than providing complete generalizable conclusions.

\textit{Internal Validity.}
We made study design trade-offs to balance efficiency and rigor, particularly in the rapid review, which was conducted by a single researcher. To mitigate this, all key steps (\ie the review protocol, and the inclusion and exclusion criteria) and intermediate findings (\ie primary studies, final themes, etc.) were discussed among all authors. The focus group data were coded and themed independently by both authors before merging. To reduce bias among our participants, we did not disclose the literature findings before the actual brainstorming session.

\textit{Construct Validity.}
Most of our data are text-based. As the literature use familiar terminology, we asses misinterpretation risks as quite low. Some ambiguity in interpreting the notes of our focus group participants may remain, though we mitigated this by asking clarification questions during the session and comparing with our own notes during the later analysis.

\section{Results}
\label{sec:results}
In this section we first outline the results from our literature review, \ie the meta-rapid review; followed by the results coming from the focus group.

\subsection{Meta-Rapid Review}
\begin{figure}[h]
  \centering
  \small
  \includegraphics[width=\linewidth]{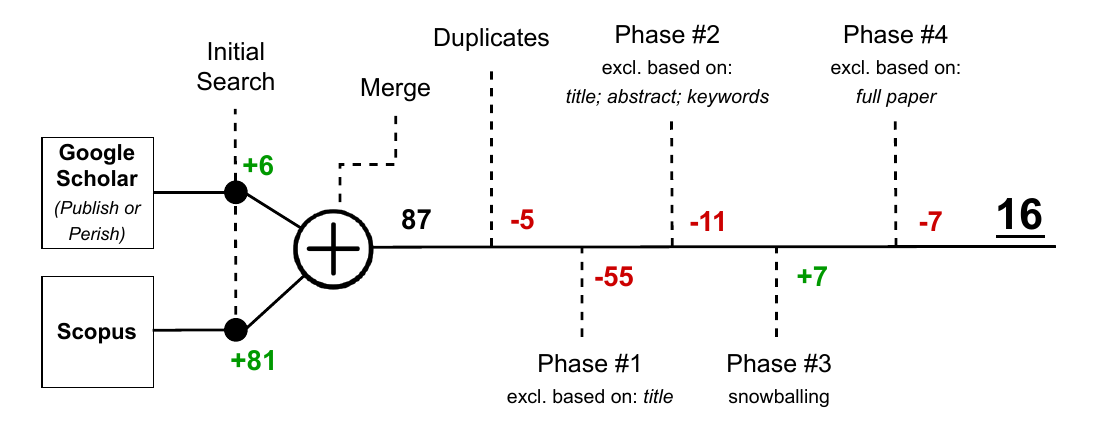}
  \caption{Meta-Rapid Review: number of publications per selection stage}
  \label{fig:RR-selection-process}
  \Description{Meta-Rapid Review: selection process and number of publications per stage.}
\end{figure}

In total, we found 16 secondary studies with the majority of the papers (n=81) resulting from Scopus (see \Cref{fig:RR-selection-process}). The studies were published between 2011 and 2025. As common, the studies target mostly certain sustainability dimensions (\citeP[P][]{Koziolek_SustainabilityEvaluation_2011b, ProcacciantiEtAl_SystematicLiterature_2015a, VerdecchiaEtAl_ArchitecturalTechnical_2018, DIasEtAl_OverviewReference_2020, BognerEtAl_IndustryPractices_2021, LiEtAl_UnderstandingSoftware_2022, DanushiEtAl_CarbonEfficientSoftware_2025, BarisicEtAl_ModellingSustainability_2025, RestrepoEtAl_SustainabledevelopmentApproach_2021, NazirEtAl_SustainableSoftware_2020}) while only 6 out of 16 (\citeP[P][]{SalamKhan_RisksMitigation_2017, AndrikopoulosEtAl_SustainabilitySoftware_2022b, FatimaLago_ReviewSoftware_2023a, AhmadisakhaAndrikopoulos_ArchitectingSustainability_2024, NakagawaKazman_WhatNew_2025, VentersEtAl_SustainableSoftware_2023a}
) acknowledge the multi-dimensional notion of sustainability and cover all dimensions.

\begin{figure}[h]
  \centering
  \includegraphics[width=\linewidth]{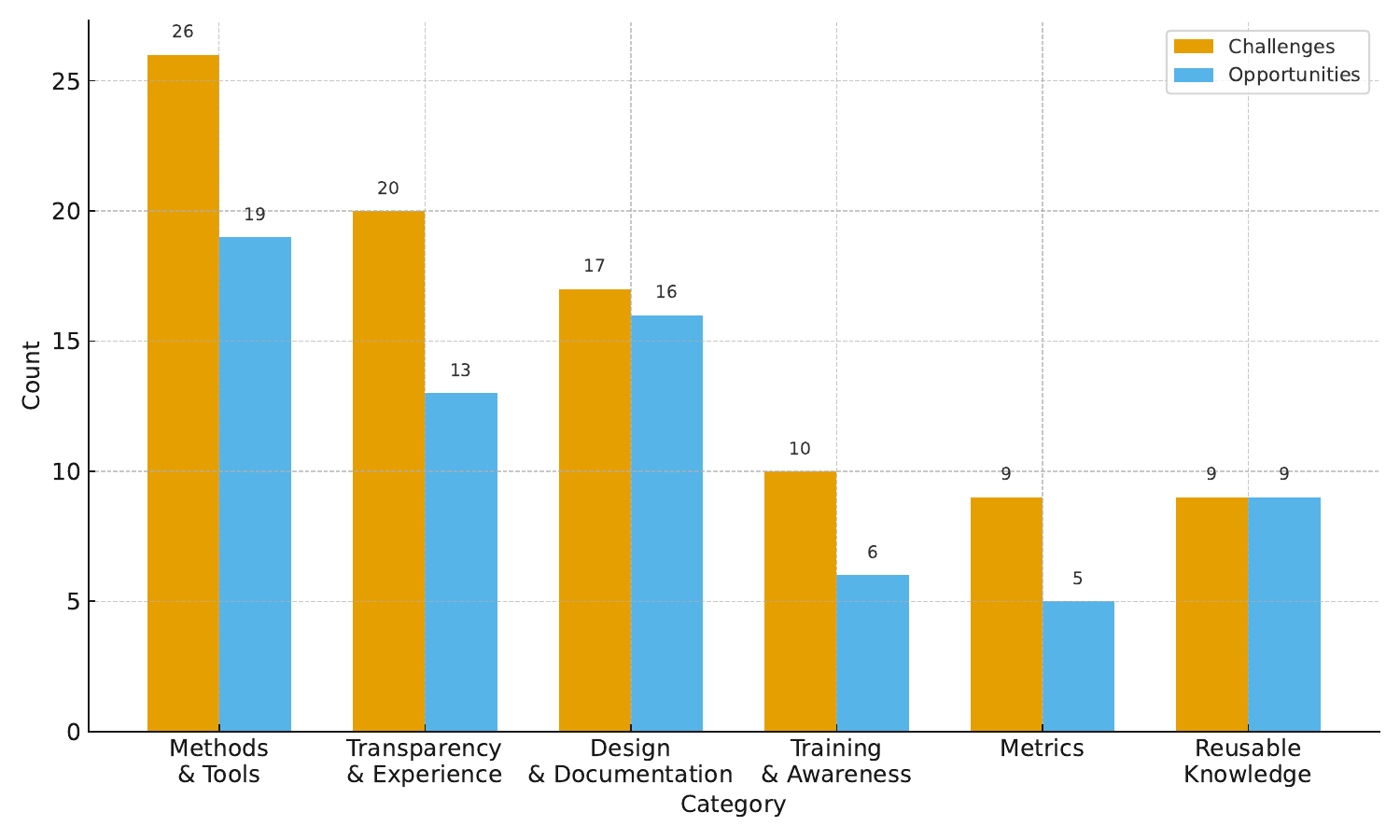}
  \caption{Meta-Rapid Review: total challenges vs. total opportunities per category (one count per paper)}
  \Description{Meta-Rapid Review: total challenges vs. total opportunities per category.}
  \label{fig:RR-graph-chall-oppor}
\end{figure}

We extracted 179 statements and categorized them into 25 sub-categories (\textit{labels}), each classified as either a \textit{challenge} or an \textit{opportunity}. These were further grouped into 6 higher-level \textit{themes}, resulting in 238 labels in total (counted at most once per paper). \Cref{fig:RR-graph-chall-oppor} presents the 6 themes with their respective counts of challenges (orange) and opportunities (blue). \Cref{tab:RR-chall-oppor} shows examples for the 7 most frequent categories ($challenge\ mentioned >= 5$).

\begin{table*}[h!]
  \caption{Meta-Rapid Review: challenges and opportunities examples per sub-category (top 7; challenge mentioned $>= 5$ times)}
  \label{tab:RR-chall-oppor}
      \small
  \begin{tabularx}{\textwidth}{p{20pt} X p{20pt} X}
    \toprule
    \textbf{num \newline chall.} & \textbf{Challenge Example} & \textbf{num \newline oppor.} & \textbf{Opportunity Example} \\
    \midrule

    \rowcolor{gray!10}\multicolumn{4}{l}{\textbf{Category:} \textit{Methods \& Tools} → \textit{Tools / Frameworks}} \\
    \rowcolor{gray!10}10 &
      ``There is a pressing need for new tooling to fit today’s emergent and dynamic environments'' \citeP[P][]{VentersEtAl_SustainableSoftware_2023a}
      & 8 &
      ``Modelling notations and tools should support iterative refinements based on the successively increasing available information during architectural design.'' \citeP[P][]{Koziolek_SustainabilityEvaluation_2011b} \\

    \rowcolor{white}\multicolumn{4}{l}{\textbf{Category:} \textit{Metrics} → \textit{Metrics}} \\
    \rowcolor{white}9 &
      ``Need for a mathematical model to quantify sustainability for all four sustainability dimensions.'' \citeP[P][]{FatimaLago_ReviewSoftware_2023a}
      & 5 &
      ``Use of energy metrics as a tool to predict the energy consumption in segments at the design stage.'' \citeP[P][]{SalamKhan_RisksMitigation_2017} \\

    \rowcolor{gray!10}\multicolumn{4}{l}{\textbf{Category:} \textit{Methods \& Tools} → \textit{SDLC}} \\
    \rowcolor{gray!10}7 &
      ``Relieving the burden from software developers [...] we should enable software architects and sustainability engineers to get more involved in the design of carbon-efficient software [...]'' \citeP[P][]{DanushiEtAl_CarbonEfficientSoftware_2025}
      & 5 &
      ``We believe that all software design activities, from requirements engineering to prototyping, can benefit from incorporating principles of sustainability-aware software design.'' \citeP[P][]{NakagawaKazman_WhatNew_2025} \\

    \rowcolor{white}\multicolumn{4}{l}{\textbf{Category:} \textit{Reusable Knowledge} → \textit{Standards / Guidelines / Patterns}} \\
    \rowcolor{white}6 &
      ``The ability to determine sustainability as a core software quality of a software system from an architectural perspective remains an open research challenge, and existing architectural principles need to be adapted and novel architectural paradigms devised.'' \citeP[P][]{VentersEtAl_SustainableSoftware_2023a}
      & 5 &
      ``Patterns and tactics are a way to encode design knowledge and make good design practice available even to inexperienced architects. Knowledge bases should be created to capture and reuse the experiences from former sustainability evaluations.'' \citeP[P][]{Koziolek_SustainabilityEvaluation_2011b} \\

    \rowcolor{gray!10}\multicolumn{4}{l}{\textbf{Category:} \textit{Design \& Documentation} → \textit{Architecture Design}} \\
    \rowcolor{gray!10}6 &
      ``New approaches are required to recover relevant views of the software architecture, including architectural design decisions, that map to architectural concepts, \ie patterns and tactics.'' \citeP[P][]{VentersEtAl_SustainableSoftware_2023a}
      & 4 &
      ``The attainment of sustainability through cloud computing becomes feasible by incorporating it into the software architecture; a concept we term sustainability through cloud.'' \citeP[P][]{AhmadisakhaAndrikopoulos_ArchitectingSustainability_2024} \\

    \rowcolor{white}\multicolumn{4}{l}{\textbf{Category:} \textit{Transparency \& Experience} → \textit{Economic Value}} \\
    \rowcolor{white}5 &
      ``Our review revealed that return on investment numbers for scenario-based methods and architecture-level metrics are currently missing.'' \citeP[P][]{Koziolek_SustainabilityEvaluation_2011b}
      & 1 &
      ``[...] help future researchers to (i) identify the economic feasibility to replicate or to use the proposed design, (ii) identify the viability in project planning and technology adoption, and (iii) know the relevance of the design to the business contexts.''~\citeP[P][]{RestrepoEtAl_SustainabledevelopmentApproach_2021} \\

    \rowcolor{gray!10}\multicolumn{4}{l}{\textbf{Category:} \textit{Training \& Awareness} → \textit{Sustainability Knowledge}} \\
    \rowcolor{gray!10}5 &
      ``Going beyond the technical dimension [...] there is an urgent need for approaches incorporating as many dimensions as possible, and addressing especially the least popular ones (environmental, economic, and social).'' \citeP[P][]{AndrikopoulosEtAl_SustainabilitySoftware_2022b}
      & 3 &
      ``Participants were very aware of the human factors of evolvability and sometimes even saw them as more challenging as technical ones. Knowledge exchange between teams was therefore a high priority for some interviewees.'' \citeP[P][]{BognerEtAl_IndustryPractices_2021} \\

    \bottomrule
  \end{tabularx}
\end{table*}

\subsubsection{Challenges}
As shown in \Cref{fig:RR-graph-chall-oppor}, \textbf{Methods \& Tools} emerged as the dominant category, reflecting a strong demand for new \textit{tools / frameworks} (n=10) to support architects in addressing sustainability. The sub-category\footnote{All sub-categories are available in the replication package and code-book online \cite{ReplicationPackage}.} \textit{SDLC} (software development life cycle) (n=7) highlights the need to integrate sustainability across the entire lifecycle, rather than limiting it to development stages. The sub-category \textit{architecture evaluation} (n=4) further indicates that current evaluation methods should be extended to cover sustainability concerns.

The second most frequently mentioned category is \textbf{Trans\-par\-en\-cy \& Experience}, reiterating the known need for greater \textit{industry involvement} (n=5) in conducting empirical studies. The literature also highlights the importance of \textit{economic value} (n=5), emphasizing the need for cost models (\eg \citeP[P][]{VentersEtAl_SustainableSoftware_2023a}) that enable evaluating trade-offs between sustainability and traditional quality concerns. Both sub-categories underline the necessity of a better cost–benefit understanding, linking sustainability impact with economic factors. 

It appears that the literature has a strong tendency towards demanding new tools and frameworks (as shown above) rather than (re-)considering existing practices in \textbf{Design \& Documentation}. In addition to challenges in \textit{architecture design} (n=6)---such as creating views that address sustainability aspects---we observe issues related to \textit{QAs} (n=3) and \textit{design decisions} (n=3). The literature still struggles to determine trade-offs between sustainability and traditional QAs, which in turn hinders informed decision-making.

The \textbf{Training \& Awareness} category was dominated by the recurring need for improved \textit{sustainability knowledge} (n=8), including the acknowledgment of the different sustainability dimensions. This gap is likely linked to insufficient \textit{training/education} (n=3) on sustainability-related topics. We were positively surprised that the \textit{temporal dimension} (n=2) was explicitly mentioned as less considered \citeP[P][]{FatimaLago_ReviewSoftware_2023a, VerdecchiaEtAl_ArchitecturalTechnical_2018}, even though it plays a key role in sustainable decision-making. The categories \textbf{Metrics} and \textbf{Reusable Knowledge} were identified with equal frequencies regarding challenges, while the demand for reusable \textit{standards/guidelines/patterns} (n=6) emerged as a major aspect within the reusable knowledge category.

\subsubsection{Opportunities}
Interestingly, as shown in \Cref{fig:RR-graph-chall-oppor}, we were not able to identify any category that contained more opportunities than challenges. Only \textbf{Reusable Knowledge} presents an equal number of opportunities and challenges. The literature recognizes potential that \textit{standards/guidelines/patterns} (n=5) and the creation of \textit{reference architectures} (n=3) could effectively have an impact on how sustainability is treated at architecture level \citeP[P][]{DIasEtAl_OverviewReference_2020}.

\textbf{Methods \& Tools} is the most dominant category mentioning concrete opportunities. \textit{Tools / frameworks} (n=8) such as a ``green gap analysis tool'' \citeP[P][]{SalamKhan_RisksMitigation_2017} or using ``neural networks'' to model sustainability \citeP[P][]{BarisicEtAl_ModellingSustainability_2025} are mentioned to solve current issues at architecture level. For \textbf{Transparency \& Experience}, \textit{stakeholder awareness} (n=4) was identified as enabler for sustainability if also stakeholders such as the end users are aware of the benefits and trade-offs of sustainable solutions. For \textbf{Design \& Documentation}, we identified that \textit{documentation} (n=3) is crucial to embed sustainability into the current way of working, leading to  \textit{architecture as enabler} (n=3), such as ``sustainability through cloud'' \citeP[P][]{AhmadisakhaAndrikopoulos_ArchitectingSustainability_2024}.

The categories \textbf{Training \& Awareness} and \textbf{Metrics} show a similar amount of opportunities, while the former recognizes the most opportunities in suggesting concrete \textit{standards/guidelines/patterns} (n=5) like ``Event Driven Messaging'' \citeP[P][]{BognerEtAl_IndustryPractices_2021} for evolvability, or ``design for reuse'' \citeP[P][]{NakagawaKazman_WhatNew_2025} for reusability; and the latter puts forward the usage of \textit{metrics} (n=5) for, \eg simulations to ``predict the energy consumption in segments at the design stage'' \citeP[P][]{SalamKhan_RisksMitigation_2017}.

\subsection{Focus Group}
Our invitation was followed up by 11 experts from 10 different companies, representing 6 distinct domains. Their roles cover various areas of expertise ranging from business consultant to solution architect. Except for 3 participants, all experts have more than 10 years of professional experience, with a median of 21 years (range: 4–36 years). The complete participant demographics is available online \cite{ReplicationPackage}.

\begin{figure}[h]
  \centering
  \label{fig:WS-graph-chall-oppor}
  \includegraphics[width=\linewidth]{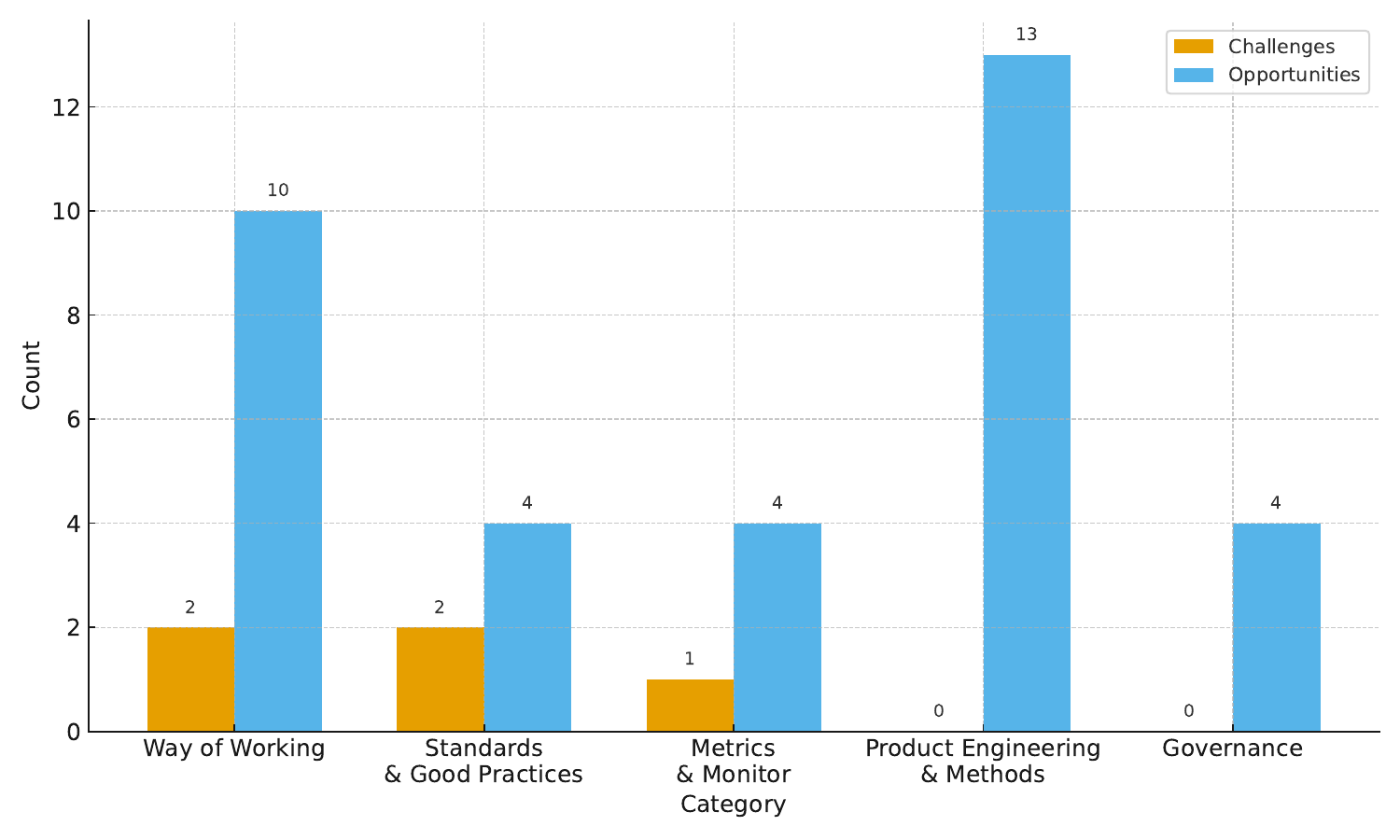}
  \caption{Focus Group: total challenges vs. total opportunities per category}
  \Description{Focus Group: total challenges vs. total opportunities per category.}
\end{figure}

During our focus group, we collected 40 notes (\ie physical sticky notes) which were classified into 5 distinct categories by both authors of this paper together. As outlined in \Cref{sec:method}, no further sub-categorization was applied due to the limited number of notes. Each note was, however, categorized as either a challenge or an opportunity by the first author of this paper, following the same approach as in the literature review.
\Cref{fig:WS-graph-chall-oppor} presents the five categories with their respective numbers of notes, divided into challenges (orange bars) and opportunities (blue bars). \Cref{tab:WS-chall-oppor} provides one concrete example of each category, mapped to the participant identifier (if applicable).

\begin{table}[h!]
  \caption{Focus Group: challenges and opportunities examples}
  \label{tab:WS-chall-oppor}
      \small
  \begin{tabularx}{\linewidth}{X X}
    \toprule
    \textbf{Challenge Example} & \textbf{Opportunity Example} \\
    \midrule

    \rowcolor{gray!10}\multicolumn{2}{l}{\textbf{Category:} \textit{Way of Working}} \\
    \rowcolor{gray!10}
      ``What wins? When sust. and some other value are diametrically opposed'' (PID-9)
      &
      ``Don't Over Engineer: Don't provision more than what is needed; Design for the near future'' (PID-1) \\

    \rowcolor{white}\multicolumn{2}{l}{\textbf{Category:} \textit{Standards \& Good Practices}} \\
    \rowcolor{white}
      ``You need guidelines of best practices which the solution architect can use as reference'' (PID-4)
      &
      ``Include it [sustainability] in the NFR [non-functional requriement] taxonomy'' (PID-10) \\
  
    \rowcolor{gray!10}\multicolumn{2}{l}{\textbf{Category:} \textit{Metrics \& Monitor}} \\
    \rowcolor{gray!10}
      ``Require energy + carbon measurements incl. monitoring'' (PID-3)
      &
      ``Think of common improvement from the start with a measuring perspective'' (n/a) \\

    \rowcolor{white}\multicolumn{2}{l}{\textbf{Category:} \textit{Product Engineering \& Methods}} \\
    \rowcolor{white}
      n/a
      &
      ``Make sure app is location independent'' (PID-7) \\
    
    \rowcolor{gray!10}\multicolumn{2}{l}{\textbf{Category:} \textit{Governance}} \\
    \rowcolor{gray!10}
      n/a
      &
      ``Include a check on sustainability aspects in the architecture governance process'' (PID-4) \\
    
    \bottomrule
  \end{tabularx}
\end{table}

\subsubsection{Challenges}
Surprisingly, we were able to categorize only 5 challenges among all notes. These challenges belong either to \textbf{Way of Working (WoW)} (n=2), \textbf{Standards \& Good Practices} (n=2), or \textbf{Metrics \& Monitor} (n=1).
Category \textbf{WoW} (n=2) reflects organizational challenges, such as questions about priorities when sustainability and other values in the process are ``diametrically opposed'' (PID-9), or in which phase sustainability (\ie ESG considerations) should be addressed.  
Category \textbf{Standards \& Good Practices} (n=2), in line with the literature, highlights the need for requirements, frameworks, and guidelines with best practices to implement sustainability at the solution architecture level.  
Finally, \textbf{Metrics \& Monitor} (n=1) points to the missing means for ``energy and carbon measurements and monitoring'' (PID-3).

\subsubsection{Opportunities}
The participants framed their notes mainly as opportunities. Category \textbf{Product Engineering \& Methods} (n=13) represents the most frequently mentioned theme, emphasizing that sustainability can be fostered through engineering at the product level, \eg through dynamic scaling, sustainable UX, location shifting, grid-aware computing, and the creation of reusable components.  

In category \textbf{WoW} (n=10), experts pointed to process-level enablers such as the need for more experience in how solution architects have actually addressed sustainability in their work, ``don’t over-engineer'' (PID-1), or ``creating a sustainability competence center to help solution architects'' (PID-4). This indicates a mindset shift from a problem focus towards a solution focus.  
Categories \textbf{Standards \& Good Practices} (n=4) and \textbf{Governance} (n=4) both stress that sustainability requires organizational commitment, \eg through strict architecture guidelines or consistent evaluation criteria that explicitly assess sustainability alongside other business standards.  
Finally, category \textbf{Metrics \& Monitor} (n=4) again emerged as an enabler, highlighting the potential of measuring energy usage to enable informed decision-making.

\section{Discussion}
\label{sec:discussion}
\begin{figure}[h]
  \centering
  \includegraphics[width=\linewidth]{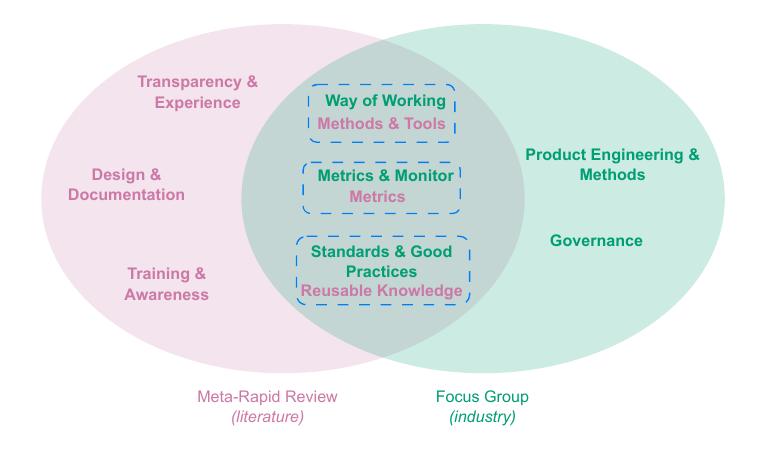}
  \caption{Overlap between findings from the meta-rapid review (literature) and focus group (industry)}
  \label{fig:RR-WS-venn}
  \Description{Overlap between findings from the meta-rapid review (literature) and focus group (industry).}
\end{figure}

Below, we synthesize findings from the meta-rapid review and the focus group to distill key takeaway messages. For consistency, we refer to them as \textit{the literature} and \textit{the industry}. We first discuss their overlaps and differences based on the Venn diagram of \Cref{fig:RR-WS-venn}, followed by key takeaways with practical implications, as typical for rapid reviews to provide actionable insights for industry \cite{CartaxoEtAl_RapidReviews_2020}.

\Cref{fig:RR-WS-venn} shows the overlap between our literature and industry findings. Since we did not reuse the literature themes for the focus group synthesis (cf. \Cref{sec:method}), a direct comparison was not feasible. However, several similar themes emerged. Such overlaps are indicated as dashed blue rectangles in the Venn diagram.

Categories such as \textit{way of working} / \textit{methods \& tools}, \textit{metrics \& monitor} / \textit{metrics}, and \textit{standards \& good practices} / \textit{reusable knowledge} are similarly represented in both sources. Interestingly, however, the industry focuses more on product-level engineering and concrete architectural decisions, while also emphasizing that sustainability needs to start earlier in the process (\eg governance). 

In contrast, the literature takes a more holistic view, addressing sustainability across the whole architecture process (cf. \textit{design \& documentation} category) and calling for stronger collaboration between academia and industry (cf. \textit{transparency \& experience})---a key point which was not discussed by our experts.

\subsection{Literature identifies Challenges while Industry finds Opportunities}
Our experts put their emphasis mostly on identifying concrete opportunities (n=35) instead of needs and challenges (n=5). In contrast, the literature categorizes more challenges (n=91) over opportunities (n=68). 
This difference might suggest that industry and academia are currently positioned at different stages. The literature mostly tries to tackle sustainability from a holistic software architecture point of view (awareness, guidelines, design decisions, etc.); instead, the industry appears to transition into an ``experimentation' phase. Curiosity and intuition driven, industry starts implementing different approaches (\eg energy efficiency tactics); while research could provide the necessary sound evidence.

\begin{tcolorbox}[boxrule=-0pt,frame hidden,sharp corners,enhanced,notitle,left=1pt,top=1pt,bottom=1pt,right=1pt,after skip=10pt plus 2pt]
    \includegraphics[height=1.2\fontcharht\font`\B]{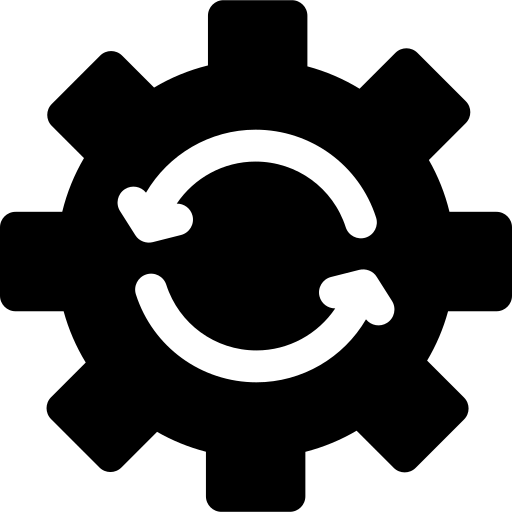} \textbf{Implication for practice:} Architects should use their intuition as starting point for their experimentation to build, keep track of, and share gained evidence with others. 
\end{tcolorbox}

\subsection{Industry is mostly concerned about Engineering Aspects focusing on the Product}
As reflected in the Venn diagram analysis, industry focuses primarily on engineering-level optimizations such as ``location shifting'' or ``dynamic scaling''. Such concrete approaches, which we would call reusable architectural tactics, were not found in the secondary studies. Only Ahmadisakha and Andrikopoulos \citeP[P][]{AhmadisakhaAndrikopoulos_ArchitectingSustainability_2024} propose ``cloud as solution'' which we would consider as an actual architectural decision. 
This shows that current initiatives tend to remain limited to the technical implementation layer. Broader aspects, such as organizational processes, governance mechanisms, or socio-economic concerns are still underexplored in practice. 

\begin{tcolorbox}[boxrule=-0pt,frame hidden,sharp corners,enhanced,notitle,left=1pt,top=1pt,bottom=1pt,right=1pt]
    \includegraphics[height=1.2\fontcharht\font`\B]{figures/insightProcess.png} \textbf{Implication for practice:} Architects should move from micro-optimizations on the product level, back to the big picture enabling long lasting impacts with the process level.
\end{tcolorbox}

\subsection{Does the Industry really need more Tools and Methods?}
Nakagawa and Kazman \citeP[P][]{NakagawaKazman_WhatNew_2025} conclude their research as follows:

\begin{quote}
    ``Factors and characteristics of sustainable software architectures are not novel [...] We believe therefore that no new factors or characteristics are necessary for sustainable architectures.'' \citeP[P][]{NakagawaKazman_WhatNew_2025}
\end{quote}

On the industry side, one expert also concludes that no new methods are necessary to address sustainability:

\begin{quote}
    ``The methods I use will not change: Application + Infra[structure], Data modeling.'' (PID-10)
\end{quote}

Although this takeaway is grounded in only one primary study and one expert, these two conclusions suggest a potentially unexplored angle: rather than developing yet another tool or framework for sustainability (currently the most common need in the literature (n=26)), we could advocate that existing architecture practices are mature enough to address sustainability as a conventional design concern. Research and practice should rather join forces to identify \textit{where} in the architecture process sustainability can be anchored by (i) reusing existing concepts (\eg guidelines, tactics, viewpoints) and (ii) adapting processes where necessary. Tool support may still aid decision-making for emerging trade-offs between sustainability and traditional quality concerns.

\begin{tcolorbox}[boxrule=-0pt,frame hidden,sharp corners,enhanced,notitle,left=1pt,top=1pt,bottom=1pt,right=1pt]
    \includegraphics[height=1.2\fontcharht\font`\B]{figures/insightProcess.png} \textbf{Implication for practice:} Rather than developing new tools, it might be worth shifting the effort towards reusing established architecture processes \textit{for} sustainability concerns.
\end{tcolorbox}

\subsection{Sustainability in Industry should become a Multi-level Architectural Concern}
Our experts were concerned with aspects going beyond concrete software or solution architectures (n=4), beginning at procurement, governance, and enterprise architecture. Governance should enable sustainability through policies and KPIs; while procurement should require sustainable criteria from suppliers. Together, these mechanisms can cascade sustainability priorities down across all architectural levels.

Similarly, both literature (n=8) and experts (n=3) suggest that sustainability should be addressed throughout the entire SDLC. \Eg Danushi \etal \citeP[P][]{DanushiEtAl_CarbonEfficientSoftware_2025} argue that we should ``avoid putting all the burden of environmental sustainability on software developers'' by trying to solve sustainability issues with coding alone; instead, efforts should support the ``design of environmentally sustainable software'' in earlier stages such as software architecture.

Together, these insights suggest that sustainability must be treated as a \textbf{multi-level architectural concern}. It needs to be addressed (i) top-down, from enterprise to implementation, and (ii) bottom-up, from development to architectural design. Unlike other traditional concerns such as availability which can often be treated `locally' at one specific level (\eg by adding redundancy or caching), sustainability is inherently multi-dimensional \cite{LagoEtAl_Toolkit_2024} and cross-cutting \cite{FunkeLago_ArchitecturalPerspective_2024}, requiring coordinated attention across all business and architectural layers. This also resembles the classic \textit{business–IT alignment} problem \cite{AversanoEtAl_LiteratureReview_2012}.

\begin{tcolorbox}[boxrule=-0pt,frame hidden,sharp corners,enhanced,notitle,left=1pt,top=1pt,bottom=1pt,right=1pt]
    \includegraphics[height=1.2\fontcharht\font`\B]{figures/insightProcess.png} \textbf{Implication for practice:} Instead of trying to solve sustainability concerns locally in one single level of architecture or development, a holistic perspective should be pursued.
\end{tcolorbox}

\subsection{Establishing a Baseline is Key before measuring or optimizing Sustainability}
One of the  most frequent challenge regards the need for sustainability metrics. Though intuitive, neither our literature nor the experts mention the need for establishing a baseline. Addressing sustainability concerns without a sound baseline risks \textit{putting the cart before the horse}. While it is indeed important to specify measures and metrics, such discussions should always go hand in hand with establishing a reliable baseline. As there is already considerable literature on energy-efficiency metrics (\eg \cite{GuldnerEtAl_DevelopmentEvaluation_2024}), the focus should shift from demanding more metrics to enabling sound monitoring.

\begin{tcolorbox}[boxrule=-0pt,frame hidden,sharp corners,enhanced,notitle,left=1pt,top=1pt,bottom=1pt,right=1pt]
    \includegraphics[height=1.2\fontcharht\font`\B]{figures/insightProcess.png} \textbf{Implication for practice:} Next to applying sound measures, a baseline is necessary to monitor the actual impacts.
\end{tcolorbox}

\section{Conclusion}
\label{sec:conclusion}
Our study investigated how to integrate and tackle sustainability concerns within software architecture by combining a rapid review of 16 secondary studies with a focus group of 11 practitioners. While the literature emphasized more challenges, experts highlighted more actionable opportunities at the product level. By combining both results, we can suggest 5 takeaway messages tailored to industry practice.
Future work should (i) investigate whether existing architectural methods are sufficient by identifying concrete \textit{hooks} in the architecting way of working; (ii) examine sustainability as a multi-level concern across all involved architecture layers; and (iii) focus more on the support for making architectural trade-offs.

%% file: 01_main.bbl

\begin{thebibliography}{32}


\ifx \showCODEN    \undefined \def \showCODEN     #1{\unskip}     \fi
\ifx \showISBNx    \undefined \def \showISBNx     #1{\unskip}     \fi
\ifx \showISBNxiii \undefined \def \showISBNxiii  #1{\unskip}     \fi
\ifx \showISSN     \undefined \def \showISSN      #1{\unskip}     \fi
\ifx \showLCCN     \undefined \def \showLCCN      #1{\unskip}     \fi
\ifx \shownote     \undefined \def \shownote      #1{#1}          \fi
\ifx \showarticletitle \undefined \def \showarticletitle #1{#1}   \fi
\ifx \showURL      \undefined \def \showURL       {\relax}        \fi
\providecommand\bibfield[2]{#2}
\providecommand\bibinfo[2]{#2}
\providecommand\natexlab[1]{#1}
\providecommand\showeprint[2][]{arXiv:#2}

\bibitem[Aversano et~al\mbox{.}(2012)]%
        {AversanoEtAl_LiteratureReview_2012}
\bibfield{author}{\bibinfo{person}{Lerina Aversano}, \bibinfo{person}{Carmine Grasso}, {and} \bibinfo{person}{Maria Tortorella}.} \bibinfo{year}{2012}\natexlab{}.
\newblock \showarticletitle{A {{Literature Review}} of {{Business}}/{{IT Alignment Strategies}}}.
\newblock \bibinfo{journal}{\emph{Procedia Technology}}  \bibinfo{volume}{5} (\bibinfo{year}{2012}).
\newblock


\bibitem[Bessin et~al\mbox{.}(2023)]%
        {bessin2023news}
\bibfield{author}{\bibinfo{person}{Anna Bessin}, \bibinfo{person}{Floris de Jong}, \bibinfo{person}{Patricia Lago}, {and} \bibinfo{person}{Oscar Widerberg}.} \bibinfo{year}{2023}\natexlab{}.
\newblock \showarticletitle{News from Europe's Digital Gateway: A Proof of Concept for Mapping Data Centre News Coverage}.
\newblock In \bibinfo{booktitle}{\emph{Environmental Informatics}}. \bibinfo{publisher}{Springer}.
\newblock


\bibitem[Calero et~al\mbox{.}(2020)]%
        {CaleroEtAl_5WsGreen_2020}
\bibfield{author}{\bibinfo{person}{Coral Calero}, \bibinfo{person}{Javier Mancebo}, \bibinfo{person}{Felix Garcia}, \bibinfo{person}{Maria~Angeles Moraga}, \bibinfo{person}{Jose Alberto~Garcia Berna}, \bibinfo{person}{Jose~Luis {Fernandez-Aleman}}, {and} \bibinfo{person}{Ambrosio Toval}.} \bibinfo{year}{2020}\natexlab{}.
\newblock \showarticletitle{{{5Ws}} of Green and Sustainable Software}.
\newblock \bibinfo{journal}{\emph{Tsinghua Science and Technology}} \bibinfo{volume}{25}, \bibinfo{number}{3} (\bibinfo{year}{2020}).
\newblock


\bibitem[Cartaxo et~al\mbox{.}(2020)]%
        {CartaxoEtAl_RapidReviews_2020}
\bibfield{author}{\bibinfo{person}{Bruno Cartaxo}, \bibinfo{person}{Gustavo Pinto}, {and} \bibinfo{person}{Sergio Soares}.} \bibinfo{year}{2020}\natexlab{}.
\newblock \showarticletitle{Rapid {{Reviews}} in {{Software Engineering}}}.
\newblock In \bibinfo{booktitle}{\emph{Contemporary {{Empirical Methods}} in {{Software Engineering}}}}. \bibinfo{publisher}{Springer International Publishing}.
\newblock


\bibitem[Cruzes and Dyba(2011)]%
        {CruzesDyba_RecommendedSteps_2011}
\bibfield{author}{\bibinfo{person}{D.~S. Cruzes} {and} \bibinfo{person}{T. Dyba}.} \bibinfo{year}{2011}\natexlab{}.
\newblock \showarticletitle{Recommended {{Steps}} for {{Thematic Synthesis}} in {{Software Engineering}}}. In \bibinfo{booktitle}{\emph{{{International Symposium}} on {{Empirical Software Engineering}} and {{Measurement}}}}. \bibinfo{publisher}{IEEE}.
\newblock


\bibitem[Desislavov et~al\mbox{.}(2023)]%
        {DesislavovEtAl_TrendsAI_2023}
\bibfield{author}{\bibinfo{person}{Radosvet Desislavov}, \bibinfo{person}{Fernando {Mart{\'i}nez-Plumed}}, {and} \bibinfo{person}{Jos{\'e} {Hern{\'a}ndez-Orallo}}.} \bibinfo{year}{2023}\natexlab{}.
\newblock \showarticletitle{Trends in {{AI}} Inference Energy Consumption: {{Beyond}} the Performance-vs-Parameter Laws of Deep Learning}.
\newblock \bibinfo{journal}{\emph{Sustainable Computing: Informatics and Systems}}  \bibinfo{volume}{38} (\bibinfo{year}{2023}).
\newblock


\bibitem[Funke and Lago(2024)]%
        {FunkeLago_ArchitecturalPerspective_2024}
\bibfield{author}{\bibinfo{person}{Markus Funke} {and} \bibinfo{person}{Patricia Lago}.} \bibinfo{year}{2024}\natexlab{}.
\newblock \showarticletitle{Towards an {{Architectural Perspective}} for {{Sustainability}}}. In \bibinfo{booktitle}{\emph{Companion {{Proceedings}} of the 15th {{International Conference}} on {{Software Business}}}}, Vol.~\bibinfo{volume}{3921}. \bibinfo{publisher}{CEUR Workshop Proceedings}.
\newblock


\bibitem[Funke and Lago(2025)]%
        {ReplicationPackage}
\bibfield{author}{\bibinfo{person}{Markus Funke} {and} \bibinfo{person}{Patricia Lago}.} \bibinfo{year}{2025}\natexlab{}.
\newblock \bibinfo{title}{Replication Package}.
\newblock
\urldef\tempurl%
\url{https://github.com/S2-group/greens-2026-rapid-review-sus-in-arch-rep-pkg}
\showURL{%
\tempurl}


\bibitem[Funke et~al\mbox{.}(2025)]%
        {FunkeEtAl_ApproachCarve_2025}
\bibfield{author}{\bibinfo{person}{Markus Funke}, \bibinfo{person}{Patricia Lago}, {and} \bibinfo{person}{Roel Donker}.} \bibinfo{year}{2025}\natexlab{}.
\newblock \showarticletitle{An Approach to Carve Sustainability into Architecture Knowledge Practice}.
\newblock \bibinfo{journal}{\emph{Information and Software Technology}}  \bibinfo{volume}{188} (\bibinfo{year}{2025}).
\newblock


\bibitem[{Garc{\'i}a-Mireles} et~al\mbox{.}(2025)]%
        {Garcia-MirelesEtAl_SustainabilityField_2025}
\bibfield{author}{\bibinfo{person}{Gabriel~Alberto {Garc{\'i}a-Mireles}}, \bibinfo{person}{Ma~{\'A}ngeles Moraga}, \bibinfo{person}{F{\'e}lix Garc{\'i}a}, {and} \bibinfo{person}{Coral Calero}.} \bibinfo{year}{2025}\natexlab{}.
\newblock \showarticletitle{Sustainability in the {{Field}} of {{Software Engineering}}: {{A Tertiary Study}}}.
\newblock \bibinfo{journal}{\emph{ACM Transactions on Software Engineering and Methodology}} (\bibinfo{year}{2025}).
\newblock


\bibitem[Groher and Weinreich(2017)]%
        {GroherWeinreich_InterviewStudy_2017}
\bibfield{author}{\bibinfo{person}{Iris Groher} {and} \bibinfo{person}{Rainer Weinreich}.} \bibinfo{year}{2017}\natexlab{}.
\newblock \showarticletitle{An {{Interview Study}} on {{Sustainability Concerns}} in {{Software Development Projects}}}. In \bibinfo{booktitle}{\emph{43rd {{Euromicro Conference}} on {{Software Engineering}} and {{Advanced Applications}} ({{SEAA}})}}. \bibinfo{publisher}{IEEE}.
\newblock


\bibitem[Gross and Ouhbi(2024)]%
        {GrossOuhbi_ClearingPath_2024}
\bibfield{author}{\bibinfo{person}{Jennifer Gross} {and} \bibinfo{person}{Sofia Ouhbi}.} \bibinfo{year}{2024}\natexlab{}.
\newblock \bibinfo{title}{Clearing the {{Path}} for {{Software Sustainability}}}.
\newblock
\urldef\tempurl%
\url{https://arxiv.org/abs/2405.15637}
\showURL{%
\tempurl}


\bibitem[Guldner et~al\mbox{.}(2024)]%
        {GuldnerEtAl_DevelopmentEvaluation_2024}
\bibfield{author}{\bibinfo{person}{Achim Guldner} {et~al\mbox{.}}} \bibinfo{year}{2024}\natexlab{}.
\newblock \showarticletitle{Development and Evaluation of a Reference Measurement Model for Assessing the Resource and Energy Efficiency of Software Products and Components---{{Green Software Measurement Model}} ({{GSMM}})}.
\newblock \bibinfo{journal}{\emph{Future Generation Computer Systems}}  \bibinfo{volume}{155} (\bibinfo{year}{2024}).
\newblock


\bibitem[Heldal et~al\mbox{.}(2024)]%
        {HeldalEtAl_SustainabilityCompetencies_2024}
\bibfield{author}{\bibinfo{person}{Rogardt Heldal}, \bibinfo{person}{Ngoc-Thanh Nguyen}, \bibinfo{person}{Ana Moreira}, \bibinfo{person}{Patricia Lago}, \bibinfo{person}{Leticia Duboc}, \bibinfo{person}{Stefanie Betz}, \bibinfo{person}{Vlad~C. Coroam{\u a}}, \bibinfo{person}{Birgit Penzenstadler}, \bibinfo{person}{Jari Porras}, \bibinfo{person}{Rafael Capilla}, \bibinfo{person}{Ian Brooks}, \bibinfo{person}{Shola Oyedeji}, {and} \bibinfo{person}{Colin~C. Venters}.} \bibinfo{year}{2024}\natexlab{}.
\newblock \showarticletitle{Sustainability Competencies and Skills in Software Engineering: {{An}} Industry Perspective}.
\newblock \bibinfo{journal}{\emph{Journal of Systems and Software}}  \bibinfo{volume}{211} (\bibinfo{year}{2024}).
\newblock


\bibitem[Karita et~al\mbox{.}(2021)]%
        {KaritaEtAl_SoftwareIndustry_2021}
\bibfield{author}{\bibinfo{person}{Leila Karita}, \bibinfo{person}{Brunna~Caroline Mour{\~a}o}, \bibinfo{person}{Luana~Almeida Martins}, \bibinfo{person}{Larissa~Rocha Soares}, {and} \bibinfo{person}{Ivan Machado}.} \bibinfo{year}{2021}\natexlab{}.
\newblock \showarticletitle{Software Industry Awareness on Sustainable Software Engineering: A {{Brazilian}} Perspective}.
\newblock \bibinfo{journal}{\emph{Journal of Software Engineering Research and Development}}  \bibinfo{volume}{9} (\bibinfo{year}{2021}).
\newblock


\bibitem[Kitchenham and Charters(2007)]%
        {KitchenhamCharters_GuidelinesPerforming_2007}
\bibfield{author}{\bibinfo{person}{Barbara Kitchenham} {and} \bibinfo{person}{Stuart Charters}.} \bibinfo{year}{2007}\natexlab{}.
\newblock \bibinfo{booktitle}{\emph{Guidelines for Performing Systematic Literature Reviews in Software Engineering}}.
\newblock \bibinfo{type}{{EBSE Technical Report}} 2.3. \bibinfo{institution}{{Keele University and University of Durham}}.
\newblock


\bibitem[Kontio et~al\mbox{.}(2008)]%
        {KontioEtAl_FocusGroup_2008}
\bibfield{author}{\bibinfo{person}{Jyrki Kontio}, \bibinfo{person}{Johanna Bragge}, {and} \bibinfo{person}{Laura Lehtola}.} \bibinfo{year}{2008}\natexlab{}.
\newblock \showarticletitle{The {{Focus Group Method}} as an {{Empirical Tool}} in {{Software Engineering}}}.
\newblock In \bibinfo{booktitle}{\emph{Guide to {{Advanced Empirical Software Engineering}}}}. \bibinfo{publisher}{Springer London}.
\newblock


\bibitem[Kumar et~al\mbox{.}(2024)]%
        {KumarEtAl_BalancingProgress_2024}
\bibfield{author}{\bibinfo{person}{Apoorva Nalini~Pradeep Kumar}, \bibinfo{person}{Justus Bogner}, \bibinfo{person}{Markus Funke}, {and} \bibinfo{person}{Patricia Lago}.} \bibinfo{year}{2024}\natexlab{}.
\newblock \showarticletitle{Balancing {{Progress}} and {{Responsibility}}: {{A Synthesis}} of {{Sustainability Trade-Offs}} of {{AI-Based Systems}}}. In \bibinfo{booktitle}{\emph{21st {{International Conference}} on {{Software Architecture Companion}} ({{ICSA-C}})}}. \bibinfo{publisher}{IEEE}.
\newblock


\bibitem[Lago et~al\mbox{.}(2024)]%
        {LagoEtAl_Toolkit_2024}
\bibfield{author}{\bibinfo{person}{Patricia Lago}, \bibinfo{person}{Nelly Condori~Fernandez}, \bibinfo{person}{Iffat Fatima}, \bibinfo{person}{Markus Funke}, {and} \bibinfo{person}{Ivano Malavolta}.} \bibinfo{year}{2024}\natexlab{}.
\newblock \showarticletitle{{The Sustainability Assessment Framework Toolkit: A Decade of Modeling Experience}}.
\newblock \bibinfo{journal}{\emph{Software \& Systems Modeling}} (\bibinfo{year}{2024}).
\newblock


\bibitem[Lammert et~al\mbox{.}(2022)]%
        {LammertEtAl_SoftwareEngineers_2022}
\bibfield{author}{\bibinfo{person}{Dominic Lammert}, \bibinfo{person}{Stefanie Betz}, {and} \bibinfo{person}{Jari Porras}.} \bibinfo{year}{2022}\natexlab{}.
\newblock \showarticletitle{Software {{Engineers}} in {{Transition}}: {{Self-Role Attribution}} and {{Awareness}} for {{Sustainability}}}. In \bibinfo{booktitle}{\emph{Hawaii {{International Conference}} on {{System Sciences}}}}.
\newblock


\bibitem[Malmodin et~al\mbox{.}(2024)]%
        {MalmodinEtAl_ICTSector_2024}
\bibfield{author}{\bibinfo{person}{Jens Malmodin}, \bibinfo{person}{Nina L{\"o}vehagen}, \bibinfo{person}{Pernilla Bergmark}, {and} \bibinfo{person}{Dag Lund{\'e}n}.} \bibinfo{year}{2024}\natexlab{}.
\newblock \showarticletitle{{{ICT}} Sector Electricity Consumption and Greenhouse Gas Emissions -- 2020 Outcome}.
\newblock \bibinfo{journal}{\emph{Telecommunications Policy}} \bibinfo{volume}{48}, \bibinfo{number}{3} (\bibinfo{year}{2024}).
\newblock


\bibitem[Manner(2023)]%
        {manner2023black}
\bibfield{author}{\bibinfo{person}{Jukka Manner}.} \bibinfo{year}{2023}\natexlab{}.
\newblock \showarticletitle{Black Software---the Energy Unsustainability of Software Systems in the 21st Century}.
\newblock \bibinfo{journal}{\emph{Oxford Open Energy}}  \bibinfo{volume}{2} (\bibinfo{year}{2023}).
\newblock


\bibitem[Motta et~al\mbox{.}(2024)]%
        {MottaEtAl_IntersectionInternet_2024}
\bibfield{author}{\bibinfo{person}{Rebeca~C. Motta}, \bibinfo{person}{Thais~V. Batista}, {and} \bibinfo{person}{Flavia~C. Delicato}.} \bibinfo{year}{2024}\natexlab{}.
\newblock \showarticletitle{The {{Intersection}} of the {{Internet}} of {{Things}} and {{Smart Cities}}: {{A Tertiary Study}}}.
\newblock \bibinfo{journal}{\emph{Journal of Internet Services and Applications}} \bibinfo{volume}{15}, \bibinfo{number}{1} (\bibinfo{year}{2024}).
\newblock


\bibitem[Peters et~al\mbox{.}(2024)]%
        {PetersEtAl_SustainabilityComputing_2024}
\bibfield{author}{\bibinfo{person}{Anne-Kathrin Peters}, \bibinfo{person}{Rafael Capilla}, \bibinfo{person}{Vlad~Constantin Coroam{\u a}}, \bibinfo{person}{Rogardt Heldal}, \bibinfo{person}{Patricia Lago}, \bibinfo{person}{Ola Leifler}, \bibinfo{person}{Ana Moreira}, \bibinfo{person}{Jo{\~a}o~Paulo Fernandes}, \bibinfo{person}{Birgit Penzenstadler}, \bibinfo{person}{Jari Porras}, {and} \bibinfo{person}{Colin~C. Venters}.} \bibinfo{year}{2024}\natexlab{}.
\newblock \showarticletitle{Sustainability in {{Computing Education}}: {{A Systematic Literature Review}}}.
\newblock \bibinfo{journal}{\emph{ACM Transactions on Computing Education}} (\bibinfo{year}{2024}).
\newblock


\bibitem[Petersen et~al\mbox{.}(2008)]%
        {PetersenEtAl_SystematicMapping_2008}
\bibfield{author}{\bibinfo{person}{Kai Petersen}, \bibinfo{person}{Robert Feldt}, \bibinfo{person}{Shahid Mujtaba}, {and} \bibinfo{person}{Michael Mattsson}.} \bibinfo{year}{2008}\natexlab{}.
\newblock \showarticletitle{Systematic {{Mapping Studies}} in {{Software Engineering}}}. In \bibinfo{booktitle}{\emph{12th {{International Conference}} on {{Evaluation}} and {{Assessment}} in {{Software Engineering}} ({{EASE}})}}.
\newblock


\bibitem[Schinckus(2020)]%
        {Schinckus_GoodBad_2020}
\bibfield{author}{\bibinfo{person}{Christophe Schinckus}.} \bibinfo{year}{2020}\natexlab{}.
\newblock \showarticletitle{The Good, the Bad and the Ugly: {{An}} Overview of the Sustainability of Blockchain Technology}.
\newblock \bibinfo{journal}{\emph{Energy Research \& Social Science}}  \bibinfo{volume}{69} (\bibinfo{year}{2020}).
\newblock


\bibitem[Trinh et~al\mbox{.}(2024)]%
        {TrinhEtAl_SustainabilityIntegration_2024}
\bibfield{author}{\bibinfo{person}{Eames Trinh}, \bibinfo{person}{Markus Funke}, \bibinfo{person}{Patricia Lago}, {and} \bibinfo{person}{Justus Bogner}.} \bibinfo{year}{2024}\natexlab{}.
\newblock \showarticletitle{Sustainability {{Integration}} of {{Artificial Intelligence}} into the {{Software Development Life Cycle}}}. In \bibinfo{booktitle}{\emph{21st {{International Conference}} on {{Software Architecture Companion}} ({{ICSA-C}})}}. \bibinfo{publisher}{IEEE}.
\newblock


\bibitem[Verdecchia et~al\mbox{.}(2023)]%
        {VerdecchiaEtAl_ThreatsValidity_2023a}
\bibfield{author}{\bibinfo{person}{Roberto Verdecchia}, \bibinfo{person}{Emelie Engstr{\"o}m}, \bibinfo{person}{Patricia Lago}, \bibinfo{person}{Per Runeson}, {and} \bibinfo{person}{Qunying Song}.} \bibinfo{year}{2023}\natexlab{}.
\newblock \showarticletitle{Threats to Validity in Software Engineering Research: {{A}} Critical Reflection}.
\newblock \bibinfo{journal}{\emph{Information and Software Technology}}  \bibinfo{volume}{164} (\bibinfo{year}{2023}).
\newblock


\bibitem[Wohlin(2014)]%
        {Wohlin_GuidelinesSnowballing_2014}
\bibfield{author}{\bibinfo{person}{Claes Wohlin}.} \bibinfo{year}{2014}\natexlab{}.
\newblock \showarticletitle{Guidelines for Snowballing in Systematic Literature Studies and a Replication in Software Engineering}. In \bibinfo{booktitle}{\emph{18th {{International Conference}} on {{Evaluation}} and {{Assessment}} in {{Software Engineering}}}}. \bibinfo{publisher}{ACM}.
\newblock


\bibitem[Wohlin et~al\mbox{.}(2012)]%
        {WohlinEtAl_ExperimentationSoftware_2012}
\bibfield{author}{\bibinfo{person}{Claes Wohlin}, \bibinfo{person}{Per Runeson}, \bibinfo{person}{Martin H{\"o}st}, \bibinfo{person}{Magnus~C. Ohlsson}, \bibinfo{person}{Bj{\"o}rn Regnell}, {and} \bibinfo{person}{Anders Wessl{\'e}n}.} \bibinfo{year}{2012}\natexlab{}.
\newblock \bibinfo{booktitle}{\emph{Experimentation in {{Software Engineering}}}}.
\newblock \bibinfo{publisher}{Springer}.
\newblock


\bibitem[{World Economic Forum}(2025)]%
        {WorldEconomicForum_2025}
\bibfield{author}{\bibinfo{person}{{World Economic Forum}}.} \bibinfo{year}{2025}\natexlab{}.
\newblock \bibinfo{booktitle}{\emph{Artificial Intelligence's Energy Paradox: {{Balancing}} Challenges and Opportunities}}.
\newblock \bibinfo{type}{White {{Paper}}}.
\newblock
\urldef\tempurl%
\url{https://reports.weforum.org/docs/WEF_Artificial_Intelligences_Energy_Paradox_2025.pdf}
\showURL{%
\tempurl}
\newblock
\shownote{accessed: 2025-10-10}.


\bibitem[Xiao(2024)]%
        {Xiao_ArchitecturalTactics_2024}
\bibfield{author}{\bibinfo{person}{Xingwen Xiao}.} \bibinfo{year}{2024}\natexlab{}.
\newblock \bibinfo{title}{Architectural {{Tactics}} to {{Improve}} the {{Environmental Sustainability}} of {{Microservices}}: {{A Rapid Review}}}.
\newblock
\urldef\tempurl%
\url{http://arxiv.org/abs/2407.16706}
\showURL{%
\tempurl}


\end{thebibliography}



\begin{thebibliography}{16}


\ifx \showCODEN    \undefined \def \showCODEN     #1{\unskip}     \fi
\ifx \showISBNx    \undefined \def \showISBNx     #1{\unskip}     \fi
\ifx \showISBNxiii \undefined \def \showISBNxiii  #1{\unskip}     \fi
\ifx \showISSN     \undefined \def \showISSN      #1{\unskip}     \fi
\ifx \showLCCN     \undefined \def \showLCCN      #1{\unskip}     \fi
\ifx \shownote     \undefined \def \shownote      #1{#1}          \fi
\ifx \showarticletitle \undefined \def \showarticletitle #1{#1}   \fi
\ifx \showURL      \undefined \def \showURL       {\relax}        \fi
\providecommand\bibfield[2]{#2}
\providecommand\bibinfo[2]{#2}
\providecommand\natexlab[1]{#1}
\providecommand\showeprint[2][]{arXiv:#2}

\bibitem[Ahmadisakha and Andrikopoulos(2024)]%
        {AhmadisakhaAndrikopoulos_ArchitectingSustainability_2024}
\bibfield{author}{\bibinfo{person}{Sahar Ahmadisakha} {and} \bibinfo{person}{Vasilios Andrikopoulos}.} \bibinfo{year}{2024}\natexlab{}.
\newblock \showarticletitle{Architecting for Sustainability of and in the Cloud: {{A}} Systematic Literature Review}.
\newblock \bibinfo{journal}{\emph{Information and Software Technology}}  \bibinfo{volume}{171} (\bibinfo{year}{2024}).
\newblock


\bibitem[Andrikopoulos et~al\mbox{.}(2022)]%
        {AndrikopoulosEtAl_SustainabilitySoftware_2022b}
\bibfield{author}{\bibinfo{person}{Vasilios Andrikopoulos}, \bibinfo{person}{Rares-Dorian Boza}, \bibinfo{person}{Carlos Perales}, {and} \bibinfo{person}{Patricia Lago}.} \bibinfo{year}{2022}\natexlab{}.
\newblock \showarticletitle{Sustainability in {{Software Architecture}}: {{A Systematic Mapping Study}}}. In \bibinfo{booktitle}{\emph{48th {{Conference}} on {{Software Engineering}} and {{Advanced Applications}} ({{SEAA}})}}. \bibinfo{publisher}{IEEE}.
\newblock


\bibitem[Bari{\v s}i{\'c} et~al\mbox{.}(2025)]%
        {BarisicEtAl_ModellingSustainability_2025}
\bibfield{author}{\bibinfo{person}{Ankica Bari{\v s}i{\'c}}, \bibinfo{person}{J{\'a}come Cunha}, \bibinfo{person}{Ivan Ruchkin}, \bibinfo{person}{Ana Moreira}, \bibinfo{person}{Jo{\~a}o Ara{\'u}jo}, \bibinfo{person}{Moharram Challenger}, \bibinfo{person}{Du{\v s}an Savi{\'c}}, {and} \bibinfo{person}{Vasco Amaral}.} \bibinfo{year}{2025}\natexlab{}.
\newblock \showarticletitle{Modelling Sustainability in Cyber--Physical Systems: {{A}} Systematic Mapping Study}.
\newblock \bibinfo{journal}{\emph{Sustainable Computing: Informatics and Systems}}  \bibinfo{volume}{45} (\bibinfo{year}{2025}).
\newblock


\bibitem[Bogner et~al\mbox{.}(2021)]%
        {BognerEtAl_IndustryPractices_2021}
\bibfield{author}{\bibinfo{person}{Justus Bogner}, \bibinfo{person}{Jonas Fritzsch}, \bibinfo{person}{Stefan Wagner}, {and} \bibinfo{person}{Alfred Zimmermann}.} \bibinfo{year}{2021}\natexlab{}.
\newblock \showarticletitle{Industry Practices and Challenges for the Evolvability Assurance of Microservices: {{An}} Interview Study and Systematic Grey Literature Review}.
\newblock \bibinfo{journal}{\emph{Empirical Software Engineering}} \bibinfo{volume}{26}, \bibinfo{number}{5} (\bibinfo{year}{2021}).
\newblock


\bibitem[Danushi et~al\mbox{.}(2025)]%
        {DanushiEtAl_CarbonEfficientSoftware_2025}
\bibfield{author}{\bibinfo{person}{Ornela Danushi}, \bibinfo{person}{Stefano Forti}, {and} \bibinfo{person}{Jacopo Soldani}.} \bibinfo{year}{2025}\natexlab{}.
\newblock \showarticletitle{Carbon-{{Efficient Software Design}} and {{Development}}: {{A Systematic Literature Review}}}.
\newblock \bibinfo{journal}{\emph{Comput. Surveys}} \bibinfo{volume}{57}, \bibinfo{number}{10} (\bibinfo{year}{2025}).
\newblock


\bibitem[Dias et~al\mbox{.}(2020)]%
        {DIasEtAl_OverviewReference_2020}
\bibfield{author}{\bibinfo{person}{Di{\'o}genes Dias}, \bibinfo{person}{Fl{\'a}via~C. Delicato}, \bibinfo{person}{Paulo~F. Pires}, \bibinfo{person}{Atslands~R. Rocha}, {and} \bibinfo{person}{Elisa~Y. Nakagawa}.} \bibinfo{year}{2020}\natexlab{}.
\newblock \showarticletitle{An Overview of Reference Architectures for Cloud of Things}. In \bibinfo{booktitle}{\emph{35th {{Annual ACM Symposium}} on {{Applied Computing}}}}. \bibinfo{publisher}{ACM}.
\newblock


\bibitem[Fatima and Lago(2023)]%
        {FatimaLago_ReviewSoftware_2023a}
\bibfield{author}{\bibinfo{person}{Iffat Fatima} {and} \bibinfo{person}{Patricia Lago}.} \bibinfo{year}{2023}\natexlab{}.
\newblock \showarticletitle{A {{Review}} of {{Software Architecture Evaluation Methods}} for {{Sustainability Assessment}}}. In \bibinfo{booktitle}{\emph{20th {{International Conference}} on {{Software Architecture Companion}} ({{ICSA-C}})}}. \bibinfo{publisher}{IEEE}.
\newblock


\bibitem[Koziolek(2011)]%
        {Koziolek_SustainabilityEvaluation_2011b}
\bibfield{author}{\bibinfo{person}{Heiko Koziolek}.} \bibinfo{year}{2011}\natexlab{}.
\newblock \showarticletitle{Sustainability Evaluation of Software Architectures: A Systematic Review}. In \bibinfo{booktitle}{\emph{Proceedings of the Joint {{ACM SIGSOFT}} Conference -- {{QoSA}} and {{ACM SIGSOFT}} Symposium -- {{ISARCS}}}}. \bibinfo{publisher}{ACM}.
\newblock


\bibitem[Li et~al\mbox{.}(2022)]%
        {LiEtAl_UnderstandingSoftware_2022}
\bibfield{author}{\bibinfo{person}{Ruiyin Li}, \bibinfo{person}{Peng Liang}, \bibinfo{person}{Mohamed Soliman}, {and} \bibinfo{person}{Paris Avgeriou}.} \bibinfo{year}{2022}\natexlab{}.
\newblock \showarticletitle{Understanding Software Architecture Erosion: {{A}} Systematic Mapping Study}.
\newblock \bibinfo{journal}{\emph{Journal of Software: Evolution and Process}} \bibinfo{volume}{34}, \bibinfo{number}{3} (\bibinfo{year}{2022}).
\newblock


\bibitem[Nakagawa and Kazman(2025)]%
        {NakagawaKazman_WhatNew_2025}
\bibfield{author}{\bibinfo{person}{Elisa~Yumi Nakagawa} {and} \bibinfo{person}{Rick Kazman}.} \bibinfo{year}{2025}\natexlab{}.
\newblock \showarticletitle{What Is {{New When Talking About Sustainable Software Architectures}}?}. In \bibinfo{booktitle}{\emph{{{IEEE}}/{{ACM International Workshop}} on {{Designing Software}} ({{Designing}})}}. \bibinfo{publisher}{IEEE}.
\newblock


\bibitem[Nazir et~al\mbox{.}(2020)]%
        {NazirEtAl_SustainableSoftware_2020}
\bibfield{author}{\bibinfo{person}{Sumaira Nazir}, \bibinfo{person}{Nargis Fatima}, \bibinfo{person}{Suriayati Chuprat}, \bibinfo{person}{Haslina Sarkan}, \bibinfo{person}{Nurulhuda F}, {and} \bibinfo{person}{Nilam~N.A. Sjarif}.} \bibinfo{year}{2020}\natexlab{}.
\newblock \showarticletitle{Sustainable {{Software Engineering}}: {{A Perspective}} of {{Individual Sustainability}}}.
\newblock \bibinfo{journal}{\emph{International Journal on Advanced Science, Engineering and Information Technology}} \bibinfo{volume}{10}, \bibinfo{number}{2} (\bibinfo{year}{2020}).
\newblock


\bibitem[Procaccianti et~al\mbox{.}(2015)]%
        {ProcacciantiEtAl_SystematicLiterature_2015a}
\bibfield{author}{\bibinfo{person}{Giuseppe Procaccianti}, \bibinfo{person}{Patricia Lago}, {and} \bibinfo{person}{Stefano Bevini}.} \bibinfo{year}{2015}\natexlab{}.
\newblock \showarticletitle{A Systematic Literature Review on Energy Efficiency in Cloud Software Architectures}.
\newblock \bibinfo{journal}{\emph{Sustainable Computing: Informatics and Systems}}  \bibinfo{volume}{7} (\bibinfo{year}{2015}).
\newblock


\bibitem[Restrepo et~al\mbox{.}(2021)]%
        {RestrepoEtAl_SustainabledevelopmentApproach_2021}
\bibfield{author}{\bibinfo{person}{Luisa Restrepo}, \bibinfo{person}{Jose Aguilar}, \bibinfo{person}{Mauricio Toro}, {and} \bibinfo{person}{Elizabeth Suesc{\'u}n}.} \bibinfo{year}{2021}\natexlab{}.
\newblock \showarticletitle{A Sustainable-Development Approach for Self-Adaptive Cyber--Physical System's Life Cycle: {{A}} Systematic Mapping Study}.
\newblock \bibinfo{journal}{\emph{JSS}}  \bibinfo{volume}{180} (\bibinfo{year}{2021}).
\newblock


\bibitem[Salam and Khan(2017)]%
        {SalamKhan_RisksMitigation_2017}
\bibfield{author}{\bibinfo{person}{M. Salam} {and} \bibinfo{person}{S.U. Khan}.} \bibinfo{year}{2017}\natexlab{}.
\newblock \showarticletitle{Risks Mitigation Practices for Multi-Sourcing Vendors in Green Software Development}.
\newblock \bibinfo{journal}{\emph{Proceedings of the Pakistan Academy of Sciences: Part A}} \bibinfo{volume}{54}, \bibinfo{number}{1A} (\bibinfo{year}{2017}).
\newblock


\bibitem[Venters et~al\mbox{.}(2023)]%
        {VentersEtAl_SustainableSoftware_2023a}
\bibfield{author}{\bibinfo{person}{Colin~C. Venters}, \bibinfo{person}{Rafael Capilla}, \bibinfo{person}{Elisa~Yumi Nakagawa}, \bibinfo{person}{Stefanie Betz}, \bibinfo{person}{Birgit Penzenstadler}, \bibinfo{person}{Tom Crick}, {and} \bibinfo{person}{Ian Brooks}.} \bibinfo{year}{2023}\natexlab{}.
\newblock \showarticletitle{Sustainable Software Engineering: {{Reflections}} on Advances in Research and Practice}.
\newblock \bibinfo{journal}{\emph{Information and Software Technology}}  \bibinfo{volume}{164} (\bibinfo{year}{2023}).
\newblock


\bibitem[Verdecchia et~al\mbox{.}(2018)]%
        {VerdecchiaEtAl_ArchitecturalTechnical_2018}
\bibfield{author}{\bibinfo{person}{Roberto Verdecchia}, \bibinfo{person}{Ivano Malavolta}, {and} \bibinfo{person}{Patricia Lago}.} \bibinfo{year}{2018}\natexlab{}.
\newblock \showarticletitle{Architectural Technical Debt Identification: The Research Landscape}. In \bibinfo{booktitle}{\emph{{{International Conference}} on {{Technical Debt}}}}. \bibinfo{publisher}{ACM}.
\newblock


\end{thebibliography}
